\documentclass[reqno,11pt]{amsart}
\usepackage{amsmath,amssymb,amsthm,graphicx,a4wide,enumerate,url}
\usepackage[small,bf]{caption} 
\newcommand{\prn}[1]{\left(#1\right)}
\newcommand{\abs}[1]{\left|#1\right|}

\newcommand{\ud}[1]{\,\mathrm{d}#1}

\begin{document}
\parskip.9ex

\title[Comparison of First and Second Order Traffic Models Via Different Data Types]
{A Comparison of Data-Fitted First Order Traffic Models and Their Second Order Generalizations Via Trajectory and Sensor Data}
\author[S. Fan]{Shimao Fan}
\address[Shimao Fan]
{Department of Mathematics \\ Temple University \\ \newline
1805 North Broad Street \\ Philadelphia, PA 19122}
\email{shimao.fan@temple.edu}
\author[B. Seibold]{Benjamin Seibold}
\address[Benjamin Seibold]
{Department of Mathematics \\ Temple University \\ \newline
1805 North Broad Street \\ Philadelphia, PA 19122}
\email{seibold@temple.edu}
\urladdr{http://www.math.temple.edu/\~{}seibold}
\subjclass[2000]{35L65; 35Q91; 91B74}
\keywords{traffic model, Lighthill-Whitham-Richards, Aw-Rascle-Zhang, second order, fundamental diagram, trajectory, sensor, data}

\begin{abstract}
The Aw-Rascle-Zhang (ARZ) model can be interpreted as a generalization of the first order Lighthill-Whitham-Richards (LWR) model, possessing a family of fundamental diagram curves, rather than a single one. We investigate to which extent this generalization increases the predictive accuracy of the models. To that end, a systematic comparison of two types of data-fitted LWR models and their second order ARZ counterparts is conducted, via a version of the three-detector problem test. The parameter functions of the models are constructed using historic fundamental diagram data. The model comparisons are then carried out using time-dependent data, of two very different types: vehicle trajectory data, and single-loop sensor data. The study of these PDE models is carried out in a macroscopic sense, i.e., continuous field quantities are constructed from the discrete data, and discretization effects are kept negligibly small.
\end{abstract}

\maketitle

\section{Introduction}
\label{sec:introduction}
In this paper the accuracy of various macroscopic traffic models is investigated, by comparing the model predictions with measurement data. Specifically, we investigate to which extent second order traffic models (here: the Aw-Rascle-Zhang (ARZ) model \cite{AwRascle2000, Zhang2002}) improve the accuracy of the predictions compared to the ``first order'' Lighthill-Whitham-Richards (LWR) model \cite{LighthillWhitham1955, Richards1956}. This question is natural, because the ARZ model can be interpreted as a generalization of the LWR model (see \S\ref{sec:traffic_models}).

The comparison is carried out via a variant of the three-detector problem \cite{Daganzo1997}: on a segment of a highway, on which the traffic state is known on both ends (at all times), a traffic model is used to simulate the evolution of the traffic state on the full segment. The thus predicted traffic state is then compared with the measured traffic state at certain positions in the interior of the segment. In order to ensure that the results are not specific to one particular type of measurement/aggregation, we consider two very different data sets: vehicle trajectory data (NGSIM data set \cite{TrafficNGSIM_I80}), and single-loop sensor data (RTMC data set \cite{TrafficMnDOT}, provided by the Minnesota Department of Transportation, Mn/DOT). Both data sets are described in \S\ref{sec:description_data}.

The LWR and the ARZ model are representatives of macroscopic traffic models, i.e., they describe the evolution of field quantities that are defined continuously in space and time, such as vehicle density, velocity, and flow rate, via partial differential equations (PDE). There is a wide variety of alternative ways to describe vehicular traffic flow, such as via microscopic models (e.g., \cite{Pipes1953, Newell1961}), mesoscopic models (e.g., \cite{HermanPrigogine1971, Phillips1979, KlarWegener2000, IllnerKlarMaterne2003}), probabilistic models (e.g., \cite{AlperovichSopasakis2008}), and cellular models (e.g., \cite{NagelSchreckenberg1992}). Moreover, the LWR and the ARZ model are inviscid models, i.e., thin transition zones of strong gradients (such as upstream ends of traffic jams) are replaced by shocks, whose evolution is determined by jump conditions \cite{Evans1998}. This is in contrast to viscous models (e.g., \cite{KernerKonhauser1993, KernerKonhauser1994}) or KdV-type models (e.g., \cite{KurtzeHong1995, KomatsuSasa1995}), which model the traffic dynamics on multiple length scales through viscosity or dispersion, respectively. An overview over many of these types of models is provided in \cite{Helbing2001}.

The macroscopic description of traffic flow does not attempt to follow each individual vehicle. Instead, information is interpreted in an aggregated (over multiple lanes) and averaged (over space and time) fashion. As a consequence, small scale noise in the drivers' behavior averages out, and a description in terms of deterministic quantities is obtained. Furthermore, many field quantities used in macroscopic models are closely linked to information that is directly recorded by stationary sensors, such as flow rate. Finally, macroscopic models are related to other means of description: they can be derived as limits of mesoscopic models \cite{NelsonSopasakis1999, KlarWegener2000}; cellular models can be derived via their discretization (``cell transmission models'') \cite{Daganzo1994, Daganzo1995_2} in Eulerian variables; and microscopic models can be derived via their discretization in Lagrangian variables (see e.g.~\cite{BorscheKimathiKlar2012}). Hence, through these relations, the results obtained in this study could be of relevance to other types of traffic models as well.

In this paper, we conduct a comparison of PDE models with data, in a fashion that is as close as possible to a macroscopic description:
\begin{itemize}
\item
The discrete measurement data is ``transformed'' into field quantities that are defined continuously in space and time (specifics are described in \S\ref{subsec:data_ngsim} for vehicle trajectory data, and in \S\ref{subsec:data_rtmc} for stationary sensor data).
\item
The numerical scheme to approximate the PDE of the traffic model is solved with a discretization small enough, so that the numerical approximation error is negligibly small relative to the errors in the models and the data (see \S\ref{sec:numerical_scheme}).
\item
The deviation of the model output from the data is measured in an $L^1$ fashion, i.e., via continuous integrals over space and/or time (see \S\ref{subsec:error_metrics} for the specific error metrics).
\end{itemize}
Using this approach the PDE models are studied in their most original form, and free from discretization effects and parameters, such as cell sizes in cell transmission models.

While the model comparison is conducted via time-dependent data, the model parameter functions are determined from historic fundamental diagram data that does not contain any temporal information (see \S\ref{sec:fd_curves}). Consequently, this study is very different from online calibration techniques, such as Kalman filters \cite{Kalman1960}, which have been quite successfully applied in the area of traffic modeling in the form of nonlinear unscented \cite{MihaylovaBoelHegyi2007}, extended \cite{WangPapageorgiou2005}, and ensemble \cite{WorkBlandinTossavainenPiccoliBayen2010} filtering. Here, once the model parameters have been determined, the model remains completely fixed, and there are no ``black-box'' parameters. Instead, all parameter functions have a distinct physical interpretation in terms of traffic flow (see \S\ref{sec:fd_curves} and \S\ref{sec:traffic_models}). That being said, we are not advocating that one perform traffic forecasting only with fixed models. Instead, it is our goal to provide some insight into which traffic models could qualify as good candidates to be used in combination with online calibration techniques.

The comparison results presented in \S\ref{sec:comparison_results} focus on two versions of the first order LWR model (corresponding to two types of flux functions), and their homogeneous ARZ model generalizations. These four models are compared in terms of their predictive accuracy with respect to measurement data. In order to guarantee a meaningful test of the models, it is ensured that in both the trajectory and the sensor data sets, a wide range of congestion level is present.

\vspace{1.5em}
\section{Description of Data Sets}
\label{sec:description_data}
For this study, we use two very different sets of data. First, we consider an NGSIM \cite{TrafficNGSIM} data set, collected on 04/13/2005 on a 500 meter segment of the eastbound direction of I-80 located in Emeryville, CA. The lane configuration is shown on \cite{TrafficNGSIM_I80}. Using video cameras, the precise trajectories of all vehicles in the segment are accessible (with a temporal resolution of 0.1 seconds), in three 15-minute intervals at the onset and during rush hour: 4:00pm--4:15pm, 5:00pm--5:15pm, and 5:15pm--5:30pm. In addition, historic fundamental diagram (FD) data for the segment, obtained via loop detectors, is available \cite{TrafficNGSIM_I80}.

The freeway segment consists of six lanes. After 125 meters an on-ramp feeds further vehicles onto the road. The ramp is a nuisance, since in this study we do not attempt to consider models for merging traffic. However, the data indicates that the ramp's influence is very small: the flow rate on the ramp is less than 8\% of the flow rate on the actual freeway lanes. Moreover, the average velocity in the rightmost lane, which the on-ramp is fed onto, is within $\pm 8\%$ of the average velocity on the other lanes. Hence, we simply treat the segment as if there were no ramp. The three recorded time-intervals possess increasing levels of congestion. 

Second, we use parts of the RTMC data set \cite{TrafficMnDOT} from the year 2003, provided by the Minnesota DOT. This data set is obtained via single loop sensors, which measure the traffic volume (number of vehicles passing a sensor, $\mathcal{N}$) and occupancy (fraction of time that a vehicle is blocking the sensor, $\sigma$) at fixed positions on the highway, aggregated over intervals of $\Delta t_\text{a}$ = 30 seconds. From these quantities, the average flow rate $\bar{Q} = \mathcal{N}/t_\text{a}$ and the average density $\bar{\rho} = \sigma/\ell$ at the sensor position are calculated, where $\ell$ is the average vehicle length, here set to $\ell$ = 5m. In the case of multiple lanes (each lane has its own sensor), the resulting values of $\bar{\rho}$ and $\bar{Q}$ are added up. The average velocity is then obtained as $\bar{u} = \bar{Q}/\bar{\rho}$. Note that significant inhomogeneities in vehicle length (e.g., due to patches of trucks) might lead to inaccuracies in the reconstructions for $\bar{\rho}$ and $\bar{u}$. This common weakness of single loop sensors could be overcome by the use of double loop detectors.

For our investigations in this paper, we consider three successive sensor positions (denoted sensors 1, 2, and 3) along the southbound direction of I-35W, south of its intersection with I-94, as shown on \cite[p.~43]{TrafficRTMC_Maps}. The highway segment between sensors 1 and 3 has a length of 1.224 km, consists of four lanes, and is free from ramps. For our model comparison, we use the sensor measurement data on 74 week days (Monday--Friday) in the time from 01/01--04/16/2003. On each day, we consider the hour between 4pm and 5pm, during which the development of rush hour traffic falls. In addition, to generate the model functions, we use historic FD data from sensor 2 (only), recorded over the whole year.

In both types of data sets, various levels of traffic congestion occur during the test intervals. The study of congested traffic states is important to create a sufficiently challenging, and thus meaningful, test for the quality of traffic models. For both types of data, the historic FD data can be used directly to create the model functions and parameters, as described in \S\ref{sec:fd_curves} and \S\ref{sec:traffic_models}. In contrast, to allow a study of macroscopic traffic models via a three-detector problem setup, macroscopic field quantities must be constructed that are defined continuously in time, and possibly in space as well. The creation of the macroscopic field quantities from the semi-discrete trajectory data or aggregated sensor data is described in \S\ref{sec:data_field_quantities}.

\vspace{1.5em}
\section{Data-Fitted Fundamental Diagram Curves}
\label{sec:fd_curves}
\begin{figure}
\begin{minipage}[t]{.49\textwidth}
\includegraphics[width=\textwidth]{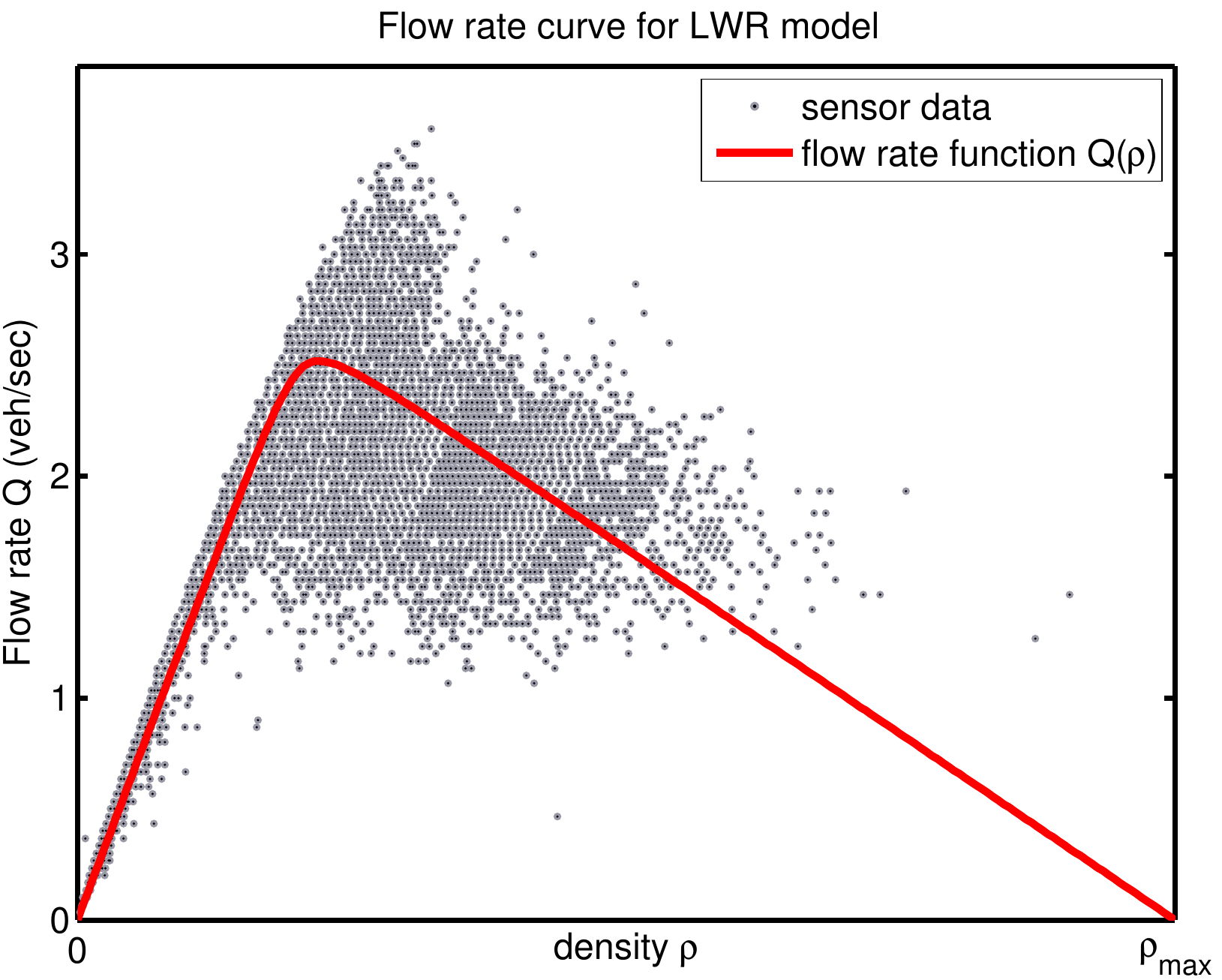}
\end{minipage}
\hfill
\begin{minipage}[t]{.49\textwidth}
\includegraphics[width=\textwidth]{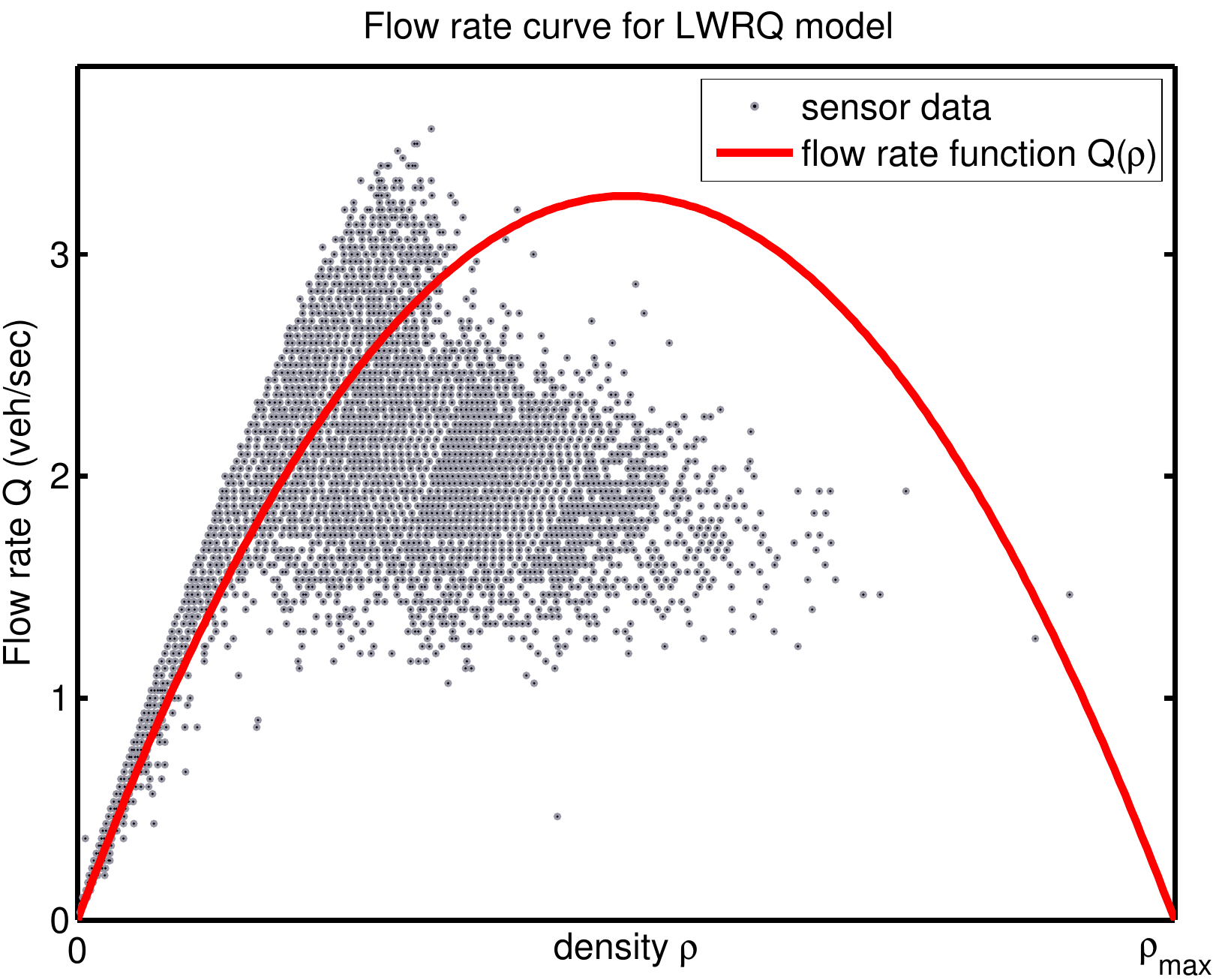}
\end{minipage}
\vspace{-.4em}
\caption{Two possible flow rate vs.\ density curve $Q(\rho)$, fitted to the historic fundamental diagram data. Left: LWR model, based on the three-parameter function \eqref{eq:flow_rate_vs_density_function}. Right: LWRQ model, using the Greenshields flux \eqref{eq:flow_rate_vs_density_function_Greenshields}.}
\label{fig:data_lwr}
\end{figure}

The fundamental diagram (FD) of traffic flow is a basic tool in traffic engineering to understand the flow capacity of a roadway. It is commonly constructed by plotting data points $(\rho,Q)$ in a flow rate vs.\ density diagram, which are obtained via long term sensor measurements. The flow rate $Q$ is the number of vehicles passing a specific position on the road per unit time, and the density $\rho$ is the average number of vehicles per unit length. Fig.~\ref{fig:data_lwr} shows a typical example of such a cloud of data points, here taken at ``sensor 2'' (see \S\ref{sec:description_data}) in the RTMC data set. One can quite distinctly see an almost perfect functional relationship between $\rho$ and $Q$ for small densities, as well as a significant spread for larger densities. Both features are typical for FDs on highways worldwide, and are commonly classified in terms of traffic phases: \emph{free flow} for low densities; and \emph{congested flow} for medium to large densities (see, e.g., \cite{Helbing2001, Varaiya2005}), or \emph{synchronized flow} for medium densities, and \emph{wide moving jams} for high densities (see Kerner \cite{Kerner1998}).

Early attempts to describe the FD of traffic flow ignored the spread in the congested flow regime, and described traffic by a ``traffic flux'' function $Q(\rho)$. The LWR model (see \S\ref{sec:traffic_models}) is defined precisely through the choice of a flux function $Q(\rho)$. A natural way to construct such a function from a measurement FD is to propose a particular functional form, and then determine the best fitting representative from the proposed class via a least squares fit to the data. While a single-valued function cannot represent the spread of the data in the congested flow regime, several of the key features of the data FD can be captured.

\subsection{Choices of Flux Functions}
A particularly simple choice of a traffic flux is the Greenshields flux \cite{Greenshields1935}
\begin{equation}
\label{eq:flow_rate_vs_density_function_Greenshields}
Q(\rho) = \rho\,u_\text{max}(1-\rho/\rho_\text{max})\;,
\end{equation}
whose quadratic form was based on Greenshields's first FD measurements conducted in the 1930s. Subsequent measurements \cite{WardropCharlesworth1954, Greenberg1959} in the 1950s illuminated that the Greenshields flux is not an accurate description of true FD data, and as a reaction many other types of traffic flux functions have been proposed and studied (see e.g.~\cite{LighthillWhitham1955, Greenberg1959, Underwood1961, Newell1993, Daganzo1994, WangPapageorgiou2005}). Specifically, in contrast to the Greenshields flux, the piecewise linear Daganzo-Newell flux function \cite{Newell1993, Daganzo1994}
\begin{equation}
\label{eq:flow_rate_vs_density_function_Daganzo_Newell}
Q(\rho) = \begin{cases}
Q_\text{max}\frac{\rho}{\rho_\text{c}} &\text{for~} 0\le\rho\le\rho_\text{c} \\
Q_\text{max}\frac{\rho_\text{max}-\rho}{\rho_\text{max}-\rho_\text{c}}
&\text{for~} \rho_\text{c}\le\rho\le\rho_\text{max}
\end{cases}
\end{equation}
captures two typical features of measured FD data: the linear relation between $Q$ and $\rho$ for small densities; and the fact that the critical density $\rho_\text{c}$, at which the flow is maximal, need not occur at $\frac{1}{2}\rho_\text{max}$. While \eqref{eq:flow_rate_vs_density_function_Daganzo_Newell} can in principle be used in the LWR model \eqref{eq:lighthill_whitham_richards_model}, it is infeasible for a second order generalization \eqref{eq:aw_rascle_zhang_model_w}, because the linear shape for $\rho\le\rho_\text{c}$ results in a loss of hyperbolicity (see \cite{FanHertySeibold2012}). Here, we ensure hyperbolicity by constructing a different class of flux functions that resemble \eqref{eq:flow_rate_vs_density_function_Daganzo_Newell}, but are smooth, strictly concave down ($Q''(\rho)<0$), and possess a strictly decreasing velocity profile ($U'(\rho) < 0$, where $U(\rho) = \frac{Q(\rho)}{\rho}$). Among the many possible choices of functions that satisfy these requirements, we use
\begin{equation}
\label{eq:flow_rate_vs_density_function}
Q(\rho) = \alpha\prn{a+(b-a)\tfrac{\rho}{\rho_{\text{max}}}-\sqrt{1+y^2}}\;,
\end{equation}
where
\begin{equation}\nonumber
a = \sqrt{1+(\lambda p)^2}\;,\quad
b = \sqrt{1+(\lambda(1-p))^2}\;,\text{~and}\quad
y = \lambda\prn{\tfrac{\rho}{\rho_{\text{max}}}-p}\;,
\end{equation}
because it has a relatively simple mathematical structure, and allows one to control the critical density $\rho_\text{c}$ (predominantly determined by $p$), the value of the maximum flow rate $Q_\text{max}$ (controlled by the factor $\alpha$), and the ``roundness'' of $Q$ (controlled by $\lambda$). Specifically, large values of $\lambda$ result in \eqref{eq:flow_rate_vs_density_function} being close to \eqref{eq:flow_rate_vs_density_function_Daganzo_Newell}, while small values of $\lambda$ yield a shape close to \eqref{eq:flow_rate_vs_density_function_Greenshields}. Note that one could equivalently write \eqref{eq:flow_rate_vs_density_function} in terms of parameters that have a direct physical meaning (such as $\rho_\text{c}$ or $Q_\text{max}$). However, the form \eqref{eq:flow_rate_vs_density_function} results in simpler expressions when conducting least squares fitting.

\subsection{Determination of Parameters}
We do not determine the stagnation density $\rho_{\text{max}}$ from data, since it is (a) generally insufficiently well represented by data, and (b) more or less determined by the physical restrictions of the road and vehicles, rather than representing drivers' behavior. We choose
\begin{equation}\nonumber
\rho_{\text{max}} = \frac{\text{\#lanes}}{7.5\text{m}}\;,
\end{equation}
where the 7.5 meters represent 5 meters average vehicle length, plus 50\% safety distance. We therefore obtain \eqref{eq:flow_rate_vs_density_function} as a three-parameter family of functions $Q_{\alpha,\lambda,p}(\rho)$.

Given historic FD measurement data $(\rho_j,Q_j),\;i=1,\dots,n$, we determine $\alpha$, $\lambda$, and $p$ by solving the LSQ minimization problem
\begin{equation}
\label{eq:LSQ_minimization}
\min_{\alpha,\lambda,p}\,\sum_{j=1}^n\abs{Q_{\alpha,\lambda,p}(\rho_j)-Q_j}^2\;.
\end{equation}
The solutions for the two data sets yield:

$\alpha = 2007\,\text{veh}/\text{sec}$, $\lambda = 16.10$, and $p = 0.189$
for the NGSIM data, and

$\alpha = 1229\,\text{veh}/\text{sec}$, $\lambda = 24.27$, and $p = 0.155$
for the RTMC data.

\noindent
As visible in the left panel of Fig.~\ref{fig:data_lwr}, the solution of \eqref{eq:LSQ_minimization} lies nicely ``in the middle'' of the cloud of FD measurement data points, and it resembles its general shape (except for the spread) quite well. In the following, this function shall be the key ingredient in the models that we denote ``LWR'' and ``ARZ''.

However, frequently the application of interest calls for even simpler functions $Q(\rho)$, for instance because rapid function evaluations are required, or because insufficient historic data is accessible. We therefore consider also the Greenshields flux \eqref{eq:flow_rate_vs_density_function_Greenshields}, which (with $\rho_{\text{max}}$ fixed) involves only a single free parameter: the empty road velocity $u_\text{max}$. Since \eqref{eq:flow_rate_vs_density_function_Greenshields} does not resemble the general shape of the FD data well (because the densities at which the flow rates become maximal do not coincide, see the right panel of Fig.~\ref{fig:data_lwr}), we do not perform a LSQ fit. Instead, $u_\text{max}$ is determined from \eqref{eq:flow_rate_vs_density_function}, such that the slopes of $Q(\rho)$ at $\rho = 0$ match, i.e., $u_\text{max} = Q'(0) = Q'_{\alpha,\lambda,p}(0)$. Since \eqref{eq:flow_rate_vs_density_function_Greenshields} is a quadratic function, we denote the models that are based on it ``LWRQ'' and ``ARZQ''.

\vspace{1.5em}
\section{First and Second Order Traffic Models}
\label{sec:traffic_models}
The basis of macroscopic traffic models is the conservation of vehicles, given by the continuity equation
\begin{equation}
\label{eq:continuity_equation}
\rho_t+(\rho u)_x = 0\;.
\end{equation}
Here $\rho(x,t)$ denotes the (lane-aggregated) density of vehicles, where $x$ is the position along a road, and $t$ is time. The Lighthill-Whitham-Richards (LWR) model \cite{LighthillWhitham1955, Richards1956} is based on a functional relationship between $\rho$ and $u$, defined by a (decreasing) velocity function $u = U(\rho)$. The induced flow rate function $Q(\rho) = \rho U(\rho)$ defines a unique $Q$ vs.~$\rho$ relationship in the associated fundamental diagram of traffic flow (see \S\ref{sec:fd_curves}). With this choice, \eqref{eq:continuity_equation} becomes a scalar hyperbolic conservation law
\begin{equation}
\label{eq:lighthill_whitham_richards_model}
\rho_t+(Q(\rho))_x = 0\;,
\end{equation}
and is thus denoted a \emph{first order model}. The two choices of data-fitted flux functions described in \S\ref{sec:fd_curves}, and depicted in Fig.~\ref{fig:data_lwr}, lead to the models ``LWR'' and ``LWRQ''.

\emph{Second order models} augment \eqref{eq:continuity_equation} by an evolution equation for the velocity field $u(x,t)$. The Payne-Whitham (PW) model \cite{Payne1971, Whitham1974} does so by introducing a ``traffic pressure'' in analogy to fluid dynamics. In order to overcome certain ``non-physical'' properties \cite{Daganzo1995} of the PW model, the Aw-Rascle-Zhang (ARZ) model \cite{AwRascle2000, Zhang2002} was proposed. The homogeneous ARZ model is
\begin{equation}
\label{eq:aw_rascle_zhang_model_homogeneous}
\begin{split}
\rho_t+(\rho u)_x &= 0\;, \\
(u+h(\rho))_t+u(u+h(\rho))_x &= 0\;,
\end{split}
\end{equation}
where we call $h(\rho)$ the \emph{hesitation function} (sometimes called ``pressure'' as well). We assume $h'(\rho)>0$ and $h(0) = 0$. The inhomogeneous ARZ model \cite{AwRascle2000, Greenberg2001} results by adding a relaxation term
\begin{equation}
\label{eq:aw_rascle_zhang_model_inhomogeneous}
\begin{split}
\rho_t+(\rho u)_x &= 0\;, \\
(u+h(\rho))_t+u(u+h(\rho))_x &= \tfrac{1}{\tau}\prn{U(\rho)-u}\;,
\end{split}
\end{equation}
implying that drivers adjust their actual velocity to the \emph{desired velocity} $U(\rho)$ with a \emph{relaxation time scale} $\tau$. In \eqref{eq:aw_rascle_zhang_model_homogeneous}, drivers do not adjust their driving behavior over time (most prominently, there is no mechanism that would cause an acceleration when starting with all vehicles at rest, $u=0$); but in \eqref{eq:aw_rascle_zhang_model_inhomogeneous} they do. Still, one can expect that \eqref{eq:aw_rascle_zhang_model_homogeneous} yield satisfactory predictions for fully established traffic flow whose nature does not change significantly in time. In this study, we restrict to the homogeneous case \eqref{eq:aw_rascle_zhang_model_homogeneous}; the inhomogeneous model \eqref{eq:aw_rascle_zhang_model_inhomogeneous} shall be considered in an upcoming paper \cite{FanHertySeibold2012}.

In analogy to fluid dynamics, the conservative form of \eqref{eq:aw_rascle_zhang_model_homogeneous} is understood as
\begin{equation}
\label{eq:aw_rascle_zhang_model_conservative}
\begin{split}
\rho_t+q_x &= 0\;, \\
q_t+\prn{\tfrac{q^2}{\rho}-h(\rho)q}_x &= 0\;,
\end{split}
\end{equation}
where the two conserved variables are $\rho$ and $q = \rho (u+h(\rho))$.

The ARZ model can be interpreted as a generalization of the LWR model, as follows. System \eqref{eq:aw_rascle_zhang_model_homogeneous} can be rewritten as
\begin{equation}
\label{eq:aw_rascle_zhang_model_w}
\begin{split}
\rho_t+(\rho u)_x &= 0\;, \\
w_t+uw_x &= 0\;, \\
\text{where~}u &= w-h(\rho)\;.
\end{split}
\end{equation}
This means that the quantity $w = u+h(\rho)$ is advected with the vehicle velocity field $u$. Moreover, for $\rho = 0$ we have that $w = u$. Therefore, $w$ can be interpreted as a property of each vehicle, namely its \emph{empty road velocity}. This interpretation renders the ARZ model to consist of a one-parameter family of velocity vs.~density curves, namely $u_w(\rho) = w-h(\rho)$. In turn, the LWR model consists of a single such curve, namely $u(\rho) = U(\rho)$. We choose
\begin{equation}\nonumber
h(\rho) = U(0)-U(\rho)\;.
\end{equation}
Thus the LWR curve is contained in the ARZ family of curves, and therefore the ARZ model can be interpreted as a generalization of the LWR model: it models the same velocity decrease with increasing density as LWR, but it allows for drivers with different empty road velocities.

\begin{figure}
\includegraphics[width=.25\textwidth]{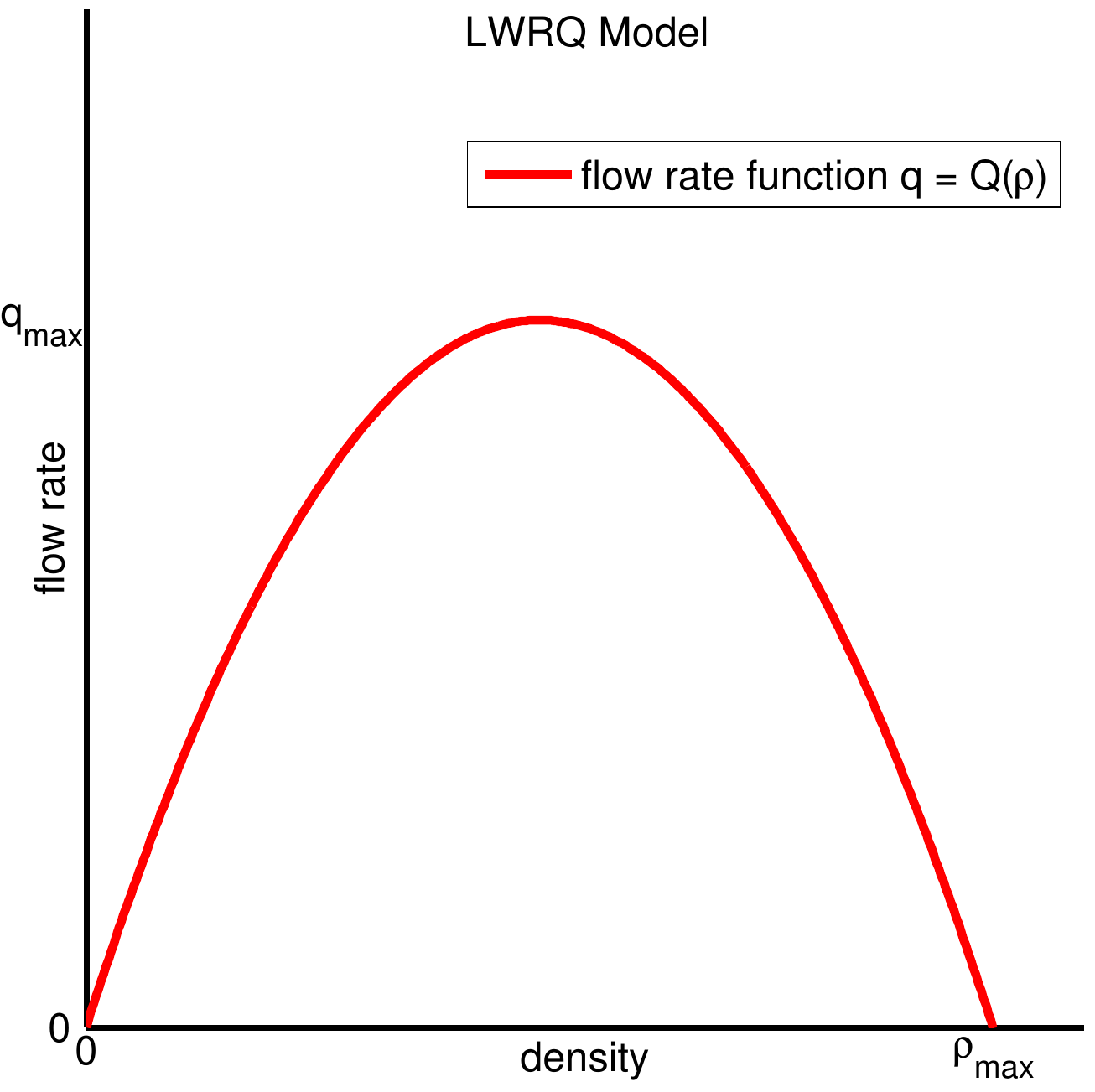}\hspace{-.008\textwidth}
\includegraphics[width=.25\textwidth]{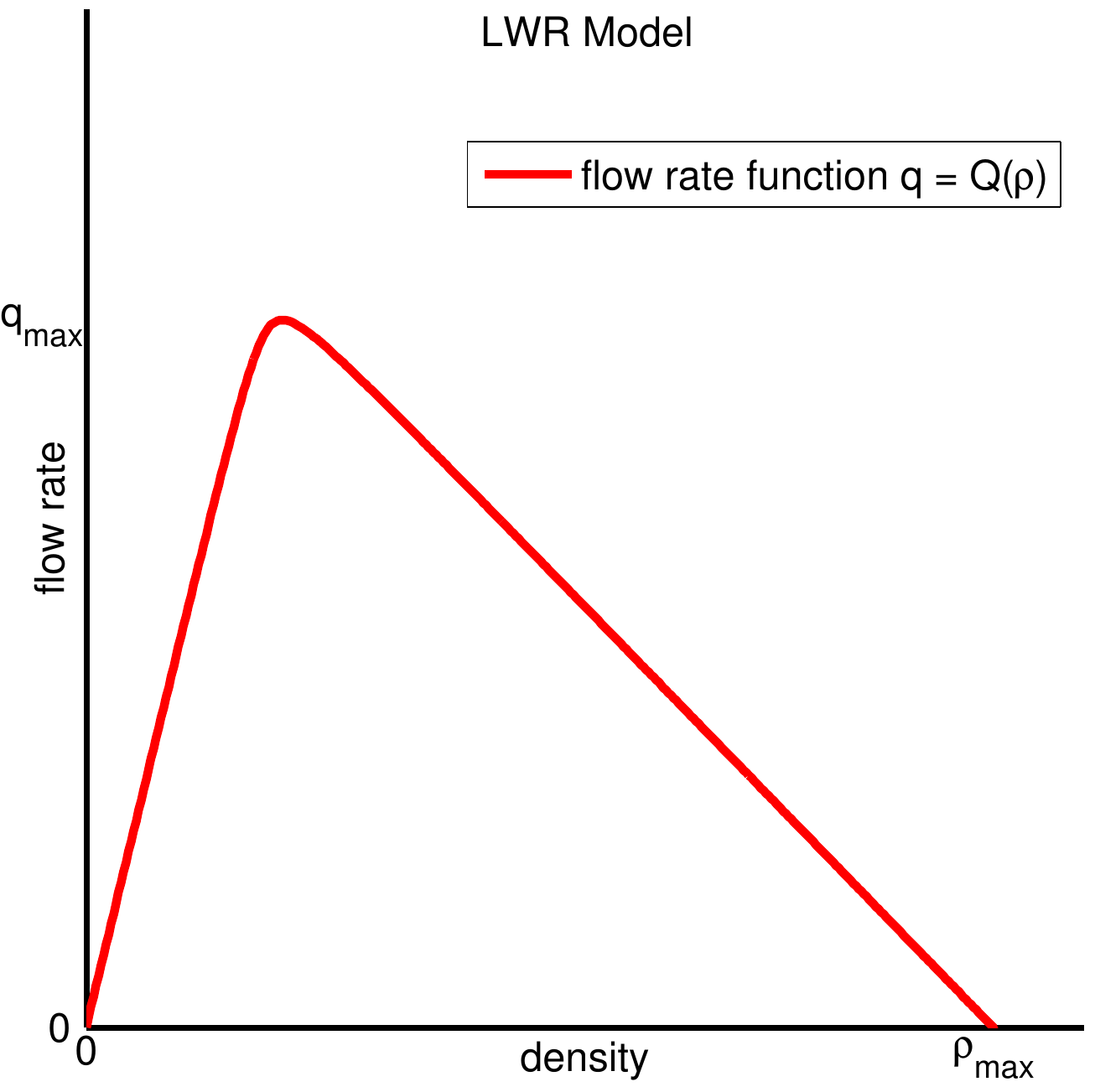}\hspace{-.008\textwidth}
\includegraphics[width=.25\textwidth]{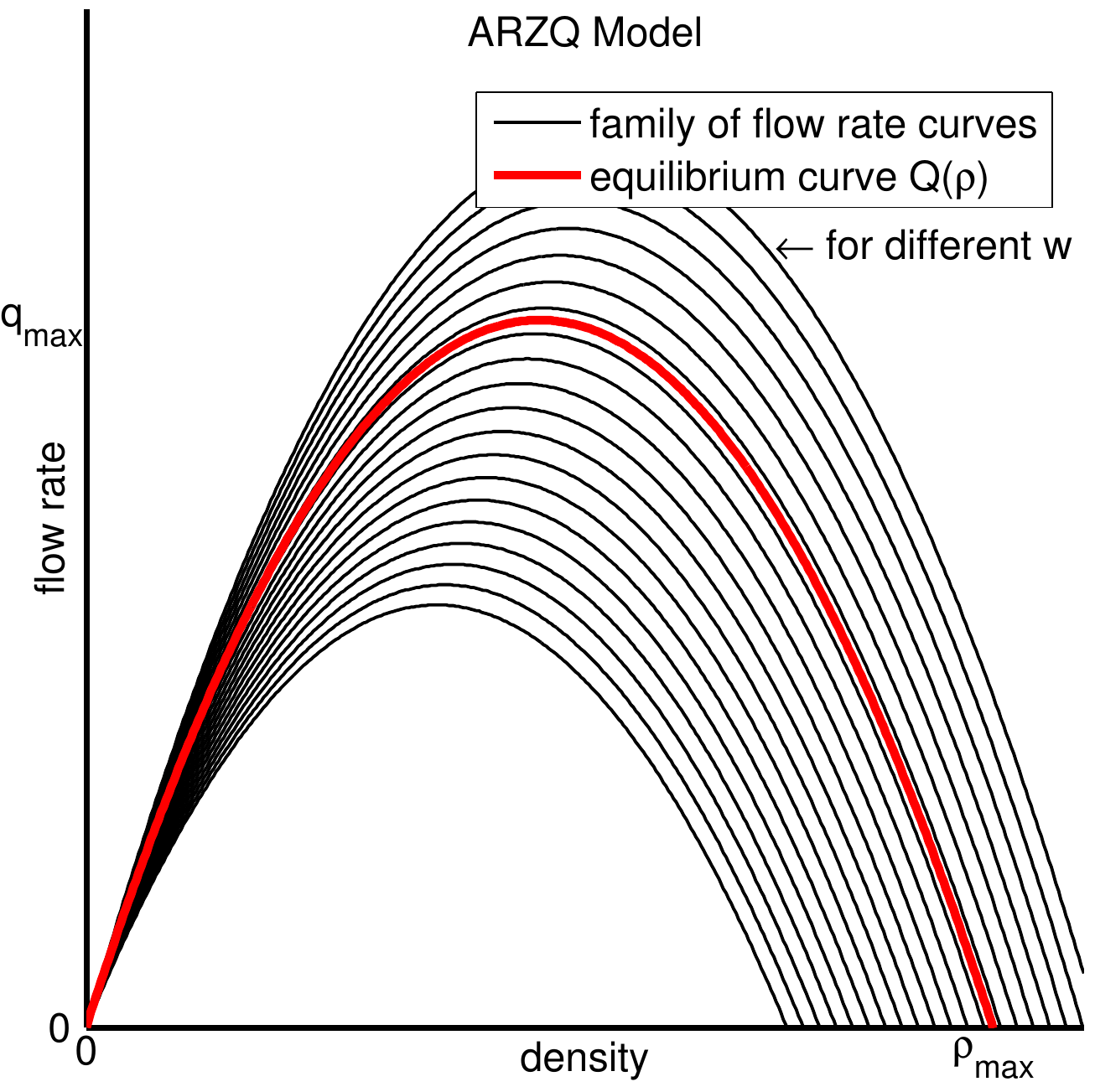}\hspace{-.008\textwidth}
\includegraphics[width=.25\textwidth]{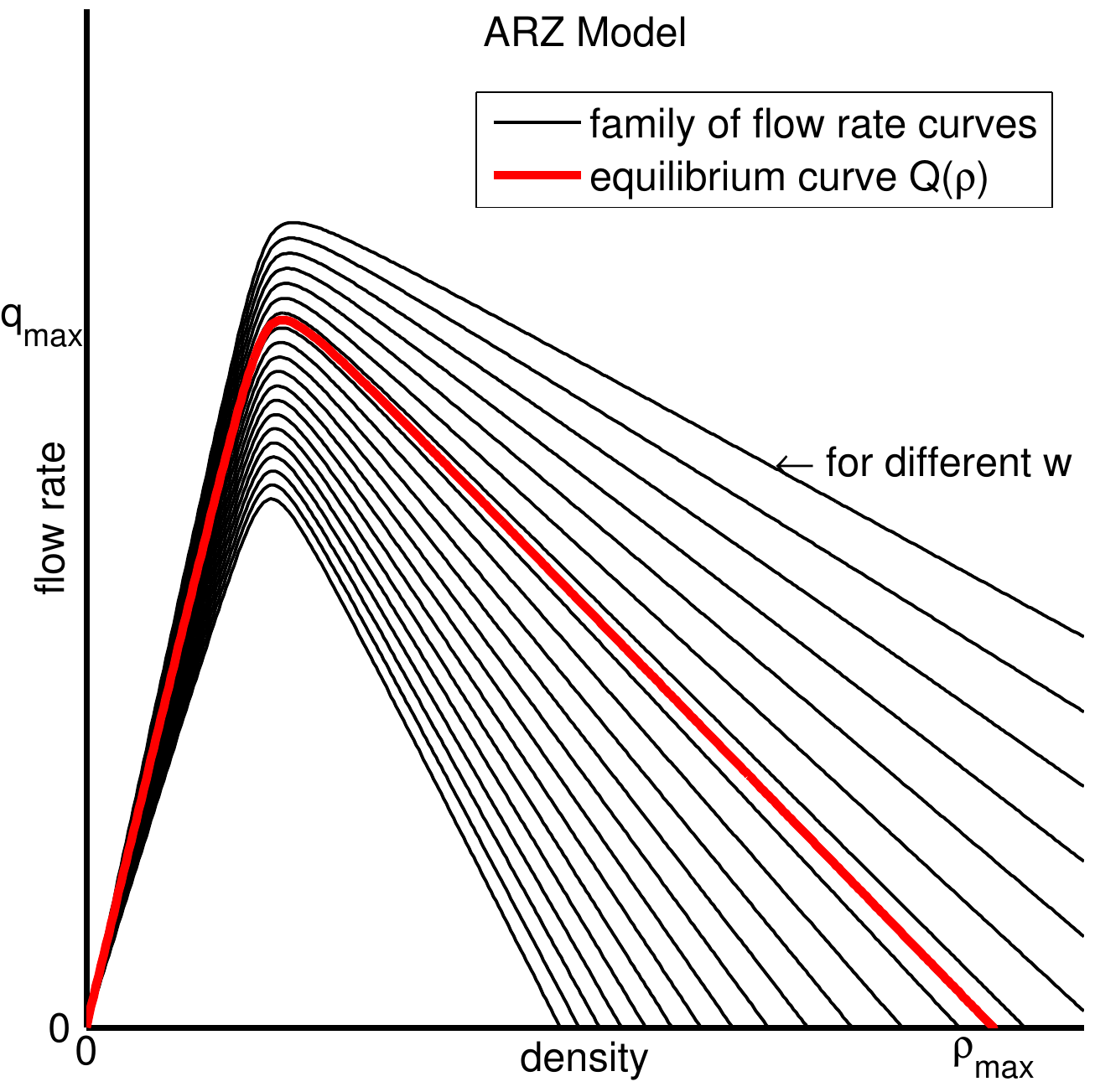}

\vspace{.7em}
\includegraphics[width=.25\textwidth]{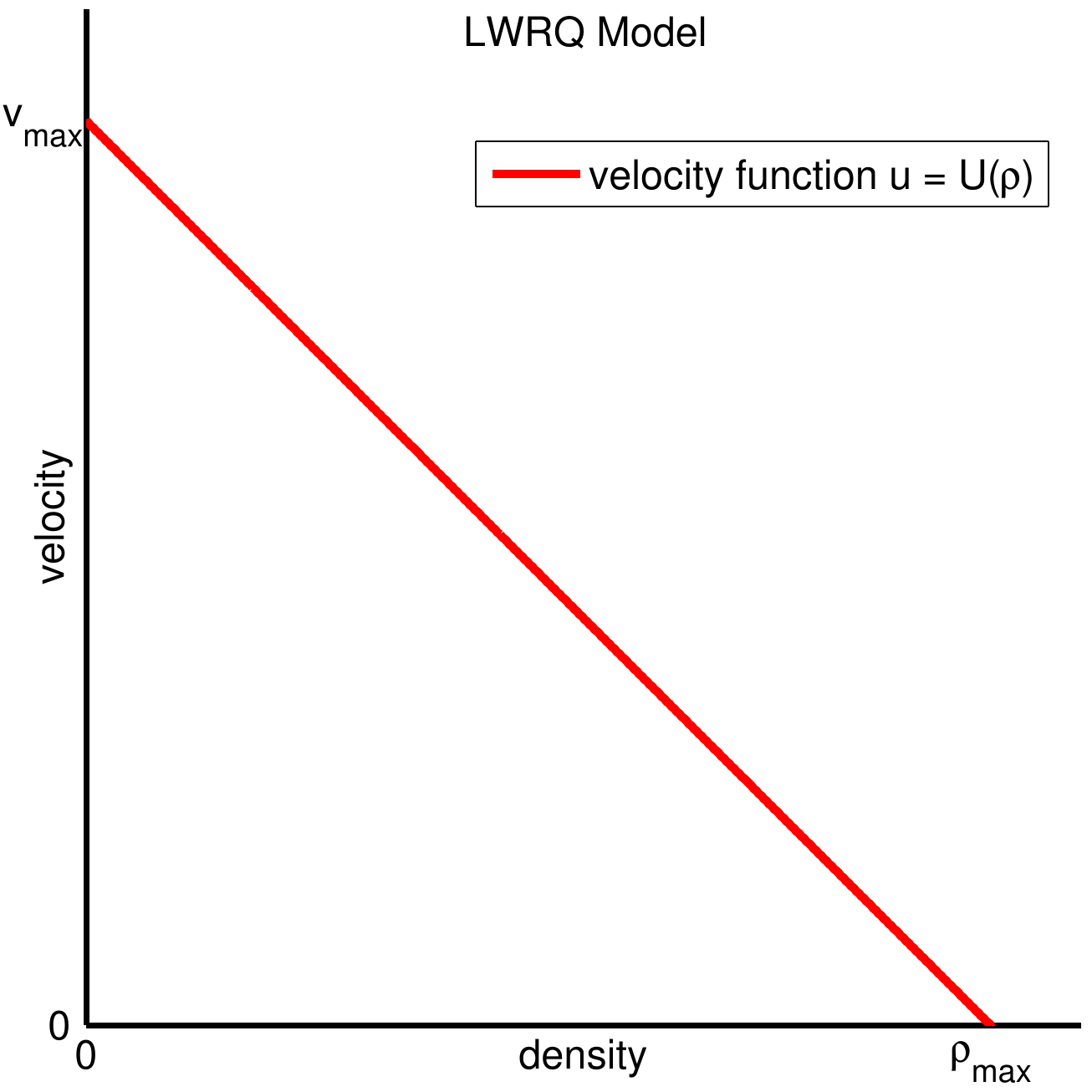}\hspace{-.008\textwidth}
\includegraphics[width=.25\textwidth]{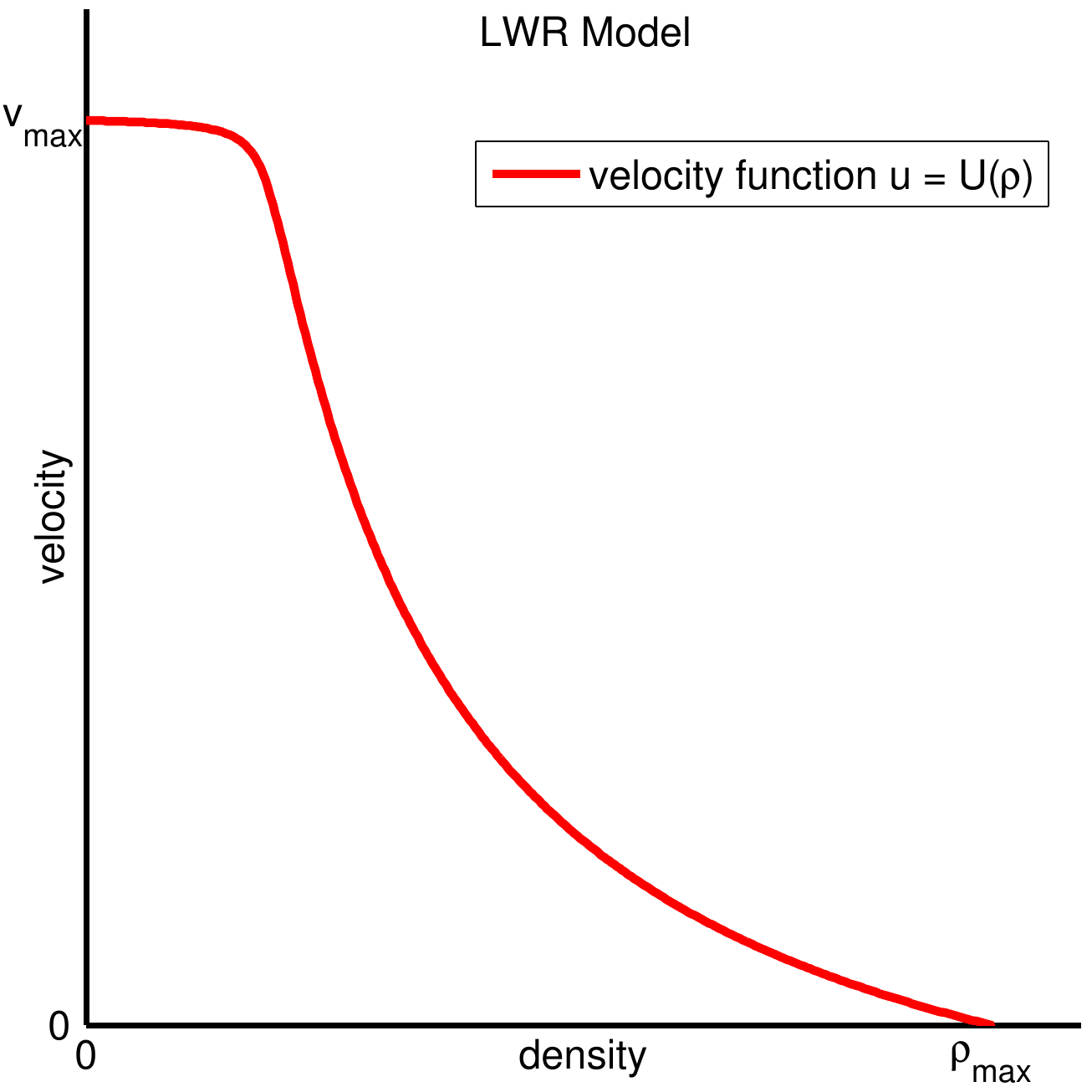}\hspace{-.008\textwidth}
\includegraphics[width=.25\textwidth]{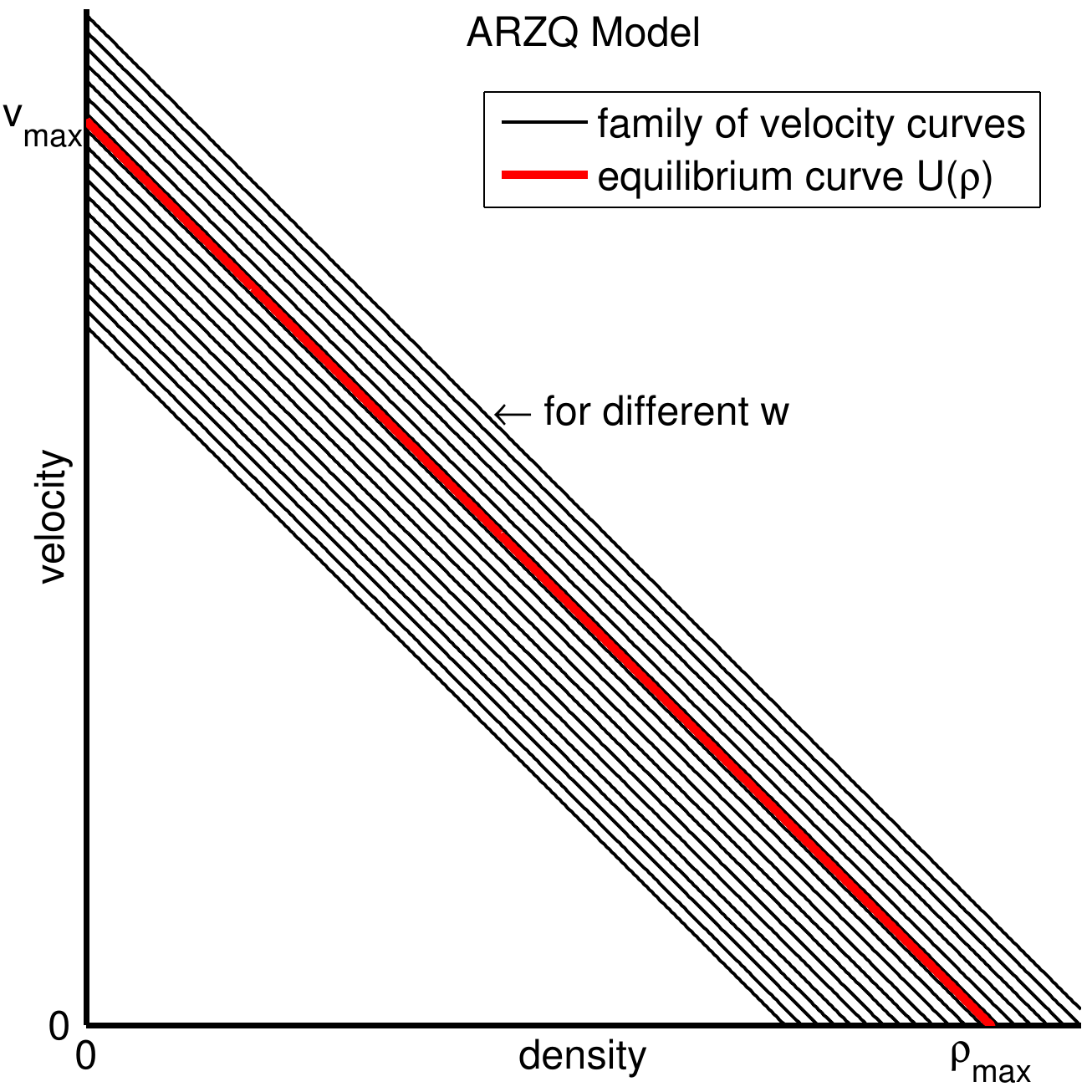}\hspace{-.008\textwidth}
\includegraphics[width=.25\textwidth]{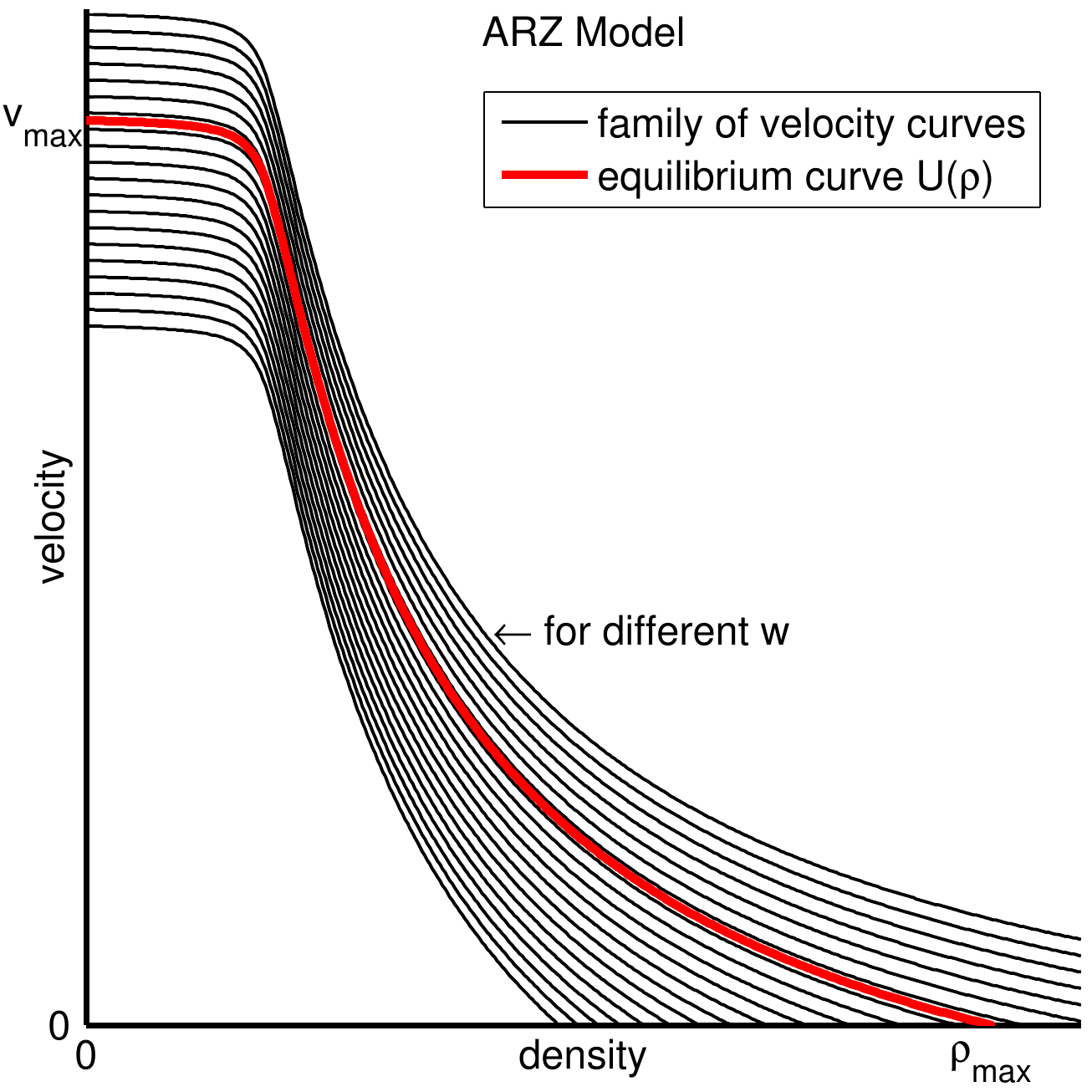}
\vspace{-1.3em}
\caption{Flow rate vs.\ density curves (top row) and corresponding velocity vs.\ density curves (bottom row) for the LWRQ, LWR, ARZQ, and ARZ models (from left to right).}
\label{fig:models}
\end{figure}

This model relationship is depicted in Fig.~\ref{fig:models}, for flow rate vs.\ density (top row), and for velocity vs.\ density (bottom row). In each row, the left two panels show the functions $Q(\rho)$ and $U(\rho)$, respectively, for the two first order models LWRQ and LWR. The right two panels in each row then correspond to the second order generalizations ARZQ and ARZ: the first order functions $Q(\rho)$ and $U(\rho)$ remain as ``equilibrium curves'' (red). In addition, a one-parameter family of curves is added, parameterized by the empty-road velocity $w$ as
\begin{equation}
\label{eq:arz_generalized_u}
u = U(\rho)+(w-U(0))\hspace{.85em}
\end{equation}
and
\begin{equation}
\label{eq:arz_generalized_Q}
Q = Q(\rho)+\rho(w-U(0))\;.
\end{equation}

Through the relationships \eqref{eq:arz_generalized_u} and \eqref{eq:arz_generalized_Q}, the second order models ARZQ and ARZ inherit the data-fitted parameter functions from their first order counterparts LWRQ and LWR. Even though the second order models do not contain any further data-fitted parameters beyond that, there are reasons to expect that they perform better at ``predicting'' the evolution of traffic flow:
\begin{itemize}
\item
In first order models, drivers adjust instantaneously to changes in density. In contrast, the ``momentum'' equation in second order models can be interpreted as a model for acceleration and deceleration. In this sense, second order models might be more realistic.
\item
At the boundaries of the study segment of road, boundary conditions must be provided. Second order models call for more boundary data than first order models, and therefore one can ``feed'' more independent data into the model (see \S\ref{subsec:boundary_conditions} for more details).
\end{itemize}

\vspace{1.5em}
\section{Numerical Scheme}
\label{sec:numerical_scheme}
The LWR model \eqref{eq:lighthill_whitham_richards_model} and the ARZ model \eqref{eq:aw_rascle_zhang_model_conservative} can both be cast as (systems of) general hyperbolic conservation laws
\begin{equation}
\label{eq:conservation_law}
\phi_t+f(\phi)_x = 0\;,
\end{equation}
for the solution $\phi(x,t)$. Here, $\phi = \rho$ and $f = Q$ for LWR, and
$\phi = (\rho,q)$ and $f(\rho,q) = (q,q^2/\rho-h(\rho)q)$ for ARZ.

\subsection{Approximation of the PDE}
We approximate the solutions of \eqref{eq:conservation_law} using a standard Godunov method \cite{Godunov1959}. This finite volume method, which is commonly used in traffic flow applications (see e.g.~\cite{LinAhanotu1995, Papageorgiou1998, JiaChenCoifmanVaraiya2001, MunozSunHorowitzAlvarez2003}), is based on the solution of local Riemann problems, which are well established for the LWR model \cite{LighthillWhitham1955} and for the ARZ model \cite{AwRascle2000}. In the numerical approach, the road segment is divided into cells of size $\Delta x$, and the solution is advanced forward in time increments $\Delta t$, which satisfy the CFL condition \cite{CourantFriedrichsLewy1928}
\begin{equation}
\label{eq:cfl_condition}
s_{\text{max}}\frac{\Delta t}{\Delta x} \leq 1\;,
\end{equation}
where $s_{\text{max}}$ is the largest wave speed. In each cell $[x_j-\frac{\Delta x}{2},x_j+\frac{\Delta x}{2}]$, the average solution value
\begin{equation}\nonumber
\phi_j^n \approx \frac{1}{\Delta x}\int_{x_{j-1/2}}^{x_{j+1/2}}\phi(x,t_n)\ud{x}
\end{equation}
is approximated. Cell-averages are updated in time by considering the balance of fluxes $f$ through the boundaries to adjacent cells. The resulting update rule reads as
\begin{equation}\nonumber
\phi_j^{n+1} = \phi_j^n-\frac{\Delta t}{\Delta x}
\prn{F_{j+\frac{1}{2}}^n-F_{j-\frac{1}{2}}^n}\;,
\end{equation}
where the Godunov flux between cell $j$ and cell $j+1$ is defined as
\begin{equation}\nonumber
F_{j+\frac{1}{2}}^n = f(\phi^*(\phi_j^n,\phi_{j+1}^n))\;,
\end{equation}
and $\phi^*(\phi_L,\phi_R)$ is the solution of the Riemann problem between the states $\phi_L$ and $\phi_R$, evaluated at $x=0$.

As argued in \S\ref{sec:introduction}, we are interested in studying the predictive accuracy of the ``pure'' PDE traffic models \eqref{eq:lighthill_whitham_richards_model} and \eqref{eq:aw_rascle_zhang_model_conservative}. To that end, we choose the spatial discretization $\Delta x$ sufficiently fine, so that the numerical approximation errors are negligibly small compared to the errors in the models and the data (we always choose $\Delta x\le 50$cm). This is in contrast to certain types of cell transmission models, in which the cell size $\Delta x$ is not small, and thus enters as an additional parameter into the model, and its influence on the solution must be considered. Note further that the Godunov method can create spurious oscillations near contact discontinuities, which can be overcome by more sophisticated schemes (e.g., \cite{ChalonsGoatin2007}). However, since here the initial and boundary data are continuous, it turns out that these oscillations are negligibly small.

\subsection{Boundary Conditions}
\label{subsec:boundary_conditions}
The hyperbolic conservation law \eqref{eq:conservation_law} describes the evolution of solutions in the interior of the space-time domain $(x,t)\in [0,L]\times [0,T]$, where $L$ is the length of the road segment, and $T$ is the final time up to which the solution is of interest. In addition, one must specify an initial condition $\phi(x,0)$, and boundary conditions. The aspect of boundary conditions for hyperbolic problems is non-trivial, since the amount and type of conditions to be specified at each boundary depends on the solution itself (see \cite{LeVeque2002}). The framework of the Godunov method allows an elegant solution to this problem: beyond each boundary (e.g., at $x=0$), a ghost cell is added (e.g., at $[-\Delta x,0]$), in which the numerical solution assumes the state of the boundary data. The Riemann solver between the outmost interior cell (e.g., $[0,\Delta x]$) and the ghost cell will then pick up exactly the right amount of data, namely the components that correspond to waves that travel into the domain $[0,L]$.

Using the measurement data, we have the density $\rho$ and the generalized flow rate $q$ available at the boundaries for all times. Since the LWR model \eqref{eq:lighthill_whitham_richards_model} is formulated in $\rho$, it uses only density boundary data (while ignoring $q$). Moreover, the boundary density only influences the result if the resulting wave is going into the domain $[0,L]$. For the ARZ model \eqref{eq:aw_rascle_zhang_model_conservative}, one wave is always right-going (a contact discontinuity that moves with the vehicle velocity, see \cite{AwRascle2000}), while the other wave could go in either direction. Specifically, for sufficiently light traffic ($\rho$ small, and $u$ large), LWR picks up boundary data for $\rho$ only at $x=0$, and nothing at $x=L$, while ARZ picks up boundary data for $\rho$ and $q$ at $x=0$, and nothing at $x=L$. In turn, for dense traffic ($\rho$ large, and $u$ small), LWR picks up boundary data for $\rho$ only at $x=L$, and nothing at $x=0$, while ARZ picks up some combination of $\rho$ and $u$ at each of the two boundaries.

\vspace{1.5em}
\section{Creation of Macroscopic Field Quantities from Measurement Data}
\label{sec:data_field_quantities}
The macroscopic traffic models \eqref{eq:lighthill_whitham_richards_model} and \eqref{eq:aw_rascle_zhang_model_conservative} require the boundary data to be defined continuously in time. The same holds true for the reference data, which is used to compare the model predictions to. And similarly, the initial data (if applicable) must be defined continuously in space. In \S\ref{subsec:data_ngsim}, we specify how continuous field quantities are constructed from the NGSIM trajectory data; and in \S\ref{subsec:data_rtmc}, the treatment of the RTMC sensor data is described.

\subsection{Treatment of NGSIM Trajectory Data}
\label{subsec:data_ngsim}
We are given the trajectories of vehicles $x_j(t)$, with a temporal resolution (0.1 seconds) that is essentially continuous in time. However, at any instance in time, the data of the vehicle positions $x_1(t),\dots,x_N(t)$, and their velocities $u_1(t),\dots,u_N(t)$, is discrete. In order to incorporate this data into, and compare it with, a PDE model, we must generate functions $\rho(x,t)$ and $u(x,t)$ that are defined everywhere on the road segment. The construction of density functions from discrete samples is an important, and well-studied, problem in statistics. As described below, we employ a kernel density estimation (KDE) approach, with a fixed Gaussian kernel, and a reflection method to remove boundary effects.

The specific KDE approach used here, also called the Parzen-Rosenblatt window method \cite{Rosenblatt1956, Parzen1962}, works as follows. Let $N$ points be given at positions $x_1,\dots,x_N$. This data is interpreted as a finite sample of some (unknown) density function $\rho(x)$. The goal is to re-construct a kernel density estimator $\hat{\rho}(x)$ from the discrete information that is close to $\rho(x)$ on an interval $[a,b]$. KDE starts with a comb function
\begin{equation}\nonumber
c(x) = \sum_{j=1}^N \delta(x-x_j)\;,
\end{equation}
and then defines $\hat{\rho}$ as its smoothed version
\begin{equation}
\label{eq:kde_density}
\hat{\rho}(x) = \int_\mathbb{R} K(x-y)c(y)\ud{y} = \sum_{j=1}^N K(x-x_j)\;,
\end{equation}
where $K$ is a smoothing kernel. Here, we consider a Gaussian kernel
\begin{equation}\nonumber
K(x) = G_h(x) = \frac{1}{\sqrt{2\pi}h}e^{-\frac{x^2}{2h^2}}\;,
\end{equation}
whose bandwidth $h$ is chosen as the smallest value so that equidistant vehicles generate an almost constant $\hat{\rho}$ for $\rho\ge\frac{1}{5}\rho_\text{max}$. For the NGSIM data sets, this amounts to $h = 25$ meters. More details on the selection of an optimal bandwidth are provided for instance in \cite{CaoCuevasManteiga1994, JonesMarronSheather1996}.

The approach can be generalized to the case in which the points carry additional information $u_1,\dots,u_N$, which is sampled from some (unknown) function $u(x)$. An estimator for the weighted density $q(x) = \rho(x)u(x)$ is obtained as
\begin{equation}
\label{eq:kde_flow_rate}
\hat{q}(x) = \int_\mathbb{R} K(x-y) \sum_{j=1}^N u_j \delta(y-x_j) \ud{y}
= \sum_{j=1}^N u_j K(x-x_j)\;.
\end{equation}
Therefore, an estimator of the function $u$ is given by
\begin{equation}
\label{eq:kde_velocity}
\hat{u}(x) = \frac{\hat{q}(x)}{\hat{\rho}(x)}
= \frac{\sum_{j=1}^N u_j K(x-x_j)}{\sum_{j=1}^N K(x-x_j)}\;.
\end{equation}
We use this approach at each time $t$ on the NGSIM data set as a unified reconstruction approach of smooth estimators for density and velocity from vehicle trajectory data. Note that the velocity function \eqref{eq:kde_velocity} is a kernel density estimator, and therefore it is in general not an interpolant of the data $u_j$ at positions $x_j$.

The estimators \eqref{eq:kde_density} and \eqref{eq:kde_flow_rate} are reasonable only at positions for which sufficiently many data points are present on either side. Therefore, a correction is required near the interval boundaries, to avoid a spurious reduction in density. We use a version of the reflection method described in \cite{KarunamuniAlberts2005}, here described for the left boundary (the right boundary is treated analogously). Let the point positions be ordered $x_1\le\dots\le x_N$, and let $d$ denote the average distance of interior points near $x_1$. Then ghost points $x_1^*,\dots,x_k^*$ are added left of $x_1$ via a reflection at $a^* = x_1-\frac{d}{2}$, i.e.
\begin{equation}\nonumber
x_j^* = 2a^*-x_j\;,
\end{equation}
and the KDE formulas \eqref{eq:kde_density} and \eqref{eq:kde_flow_rate} are applied to the thus augmented data set (left of $x_1$, and right of $x_N$). On each side, the number of ghost points $k$ is chosen, such that the estimators \eqref{eq:kde_density} and flow rate \eqref{eq:kde_flow_rate} do not ``feel'' the boundary effects anymore.

\subsection{Treatment of the RTMC Sensor Data}
\label{subsec:data_rtmc}
At the sensor positions, we reconstruct continuous density, velocity, and flow rate functions from the aggregated data of the respective quantities, as follows. For each interval $[t_n,t_n+\Delta t_\text{a}]$, we use the averaged quantity to define the value of the continuous function at the interval's mid-time, i.e., $\rho(t_n+\frac{1}{2}\Delta t_\text{a}) = \bar{\rho}_n$. Then, using these data points, the function $\rho(t)$ is defined for all times by using cubic interpolation. The procedure for $u$ and $Q$ is analogous. This time-continuous data is then used at sensors 1 and 3 to provide boundary conditions, and at sensor 2 to perform a comparison with the predictions of the model.

In addition to boundary conditions, an initial state is needed to run a forward simulation of a macroscopic traffic model. Unfortunately, a state estimation from stationary sensors that are relatively far apart is not possible with a level of accuracy that would be required here, i.e., the model comparison would not be overshadowed by errors due to the state estimation. Therefore, we choose an alternative approach: before the comparison between traffic model and data is conducted, we run the traffic model through an initialization phase (about 5 minutes long), which is started with a uniform, low density traffic state. During this initialization phase, as information flows into the domain from the correct boundary data, the traffic model ``fills'' the interval with a reasonable state.

\vspace{1.5em}
\section{Comparison Results}
\label{sec:comparison_results}
We now apply the four data-fitted traffic models LWRQ, LWR, ARZQ, and ARZ to the different sets of time-dependent boundary data to generate predictions of the evolution of the traffic state, which we then compare with time-dependent reference data in the interior of the domain. In \S\ref{subsec:error_metrics}, we define precise metrics that quantify the difference of a model prediction with measurement data. Then, the results of the comparisons are presented, with the NGSIM data in \S\ref{subsec:comparison_ngsim}, and the RTMC data in \S\ref{subsec:comparison_rtmc}.

\subsection{Error Metrics for Model Accuracy Quantification}
\label{subsec:error_metrics}
The models produce predictions of the traffic state in the interior of the space-time domain $(x,t)\in (0,L)\times (0,T)$. In order to quantify the predictive accuracy of a model, we evaluate its ``error'', i.e., the deviation of the predicted state from the state given by measurement data. We use an error metric of the following form. First, we assume that one is interested in predicting both traffic densities and velocities accurately, and therefore include both quantities into the error metric. Given a model output $\rho^\text{model}$ and $u^\text{model}$, and data $\rho^\text{data}$ and $u^\text{data}$ at the same position and time, we define the scaled difference
\begin{equation}
\label{eq:error_e}
E(x,t) = \frac{\abs{\rho^\text{data}(x,t)-\rho^\text{model}(x,t)}}{\rho_\text{max}}
+ \frac{\abs{u^\text{data}(x,t)-u^\text{model}(x,t)}}{u_\text{max}}\;.
\end{equation}
Second, to remove the impact of noise and outliers in the data, we consider averages (in an $L^1$ sense). Specifically, we consider spatial averages
\begin{equation}
\label{eq:error_x}
E^x(t) = \frac{1}{L}\int_0^L E(x,t) \ud{x}
\end{equation}
for the NGSIM data set, and temporal averages
\begin{equation}
\label{eq:error_t}
E^t(x) = \frac{1}{T}\int_0^T E(x,t) \ud{t}
\end{equation}
for the RTMC data set (evaluated at $x_2$, the position of sensor 2). For each of the three NGSIM data sets we plot $E^x(t)$ as a function of time, and for the RTMC data we plot the value of $E^t(x_2)$ for each of the 74 considered days.

Moreover, for each of the NGSIM data sets, we obtain a final error for each model by averaging over space and time
\begin{equation}
\label{eq:error_xt}
E = \frac{1}{TL}\int_0^T \int_0^L E(x,t) \ud{x} \ud{t}\;,
\end{equation}
and for the RTMC data, we obtain a fully averaged error for each model by averaging over all days
\begin{equation}
\label{eq:error_days}
E = \frac{1}{\text{\#days}}\sum_{j=1}^{\text{\#days}} E^t_\text{day $j$}(x_2)\;.
\end{equation}
Since the model predictions result from a numerical scheme with a (small) cell size and time step, the integrals in \eqref{eq:error_x}, \eqref{eq:error_t}, and \eqref{eq:error_xt} are approximated (with negligible approximation error) using quadrature rules over the computational space-time grid.

\subsection{Model Comparison using the NGSIM Trajectory Data}
\label{subsec:comparison_ngsim}
\begin{figure}
\includegraphics[width=\textwidth]{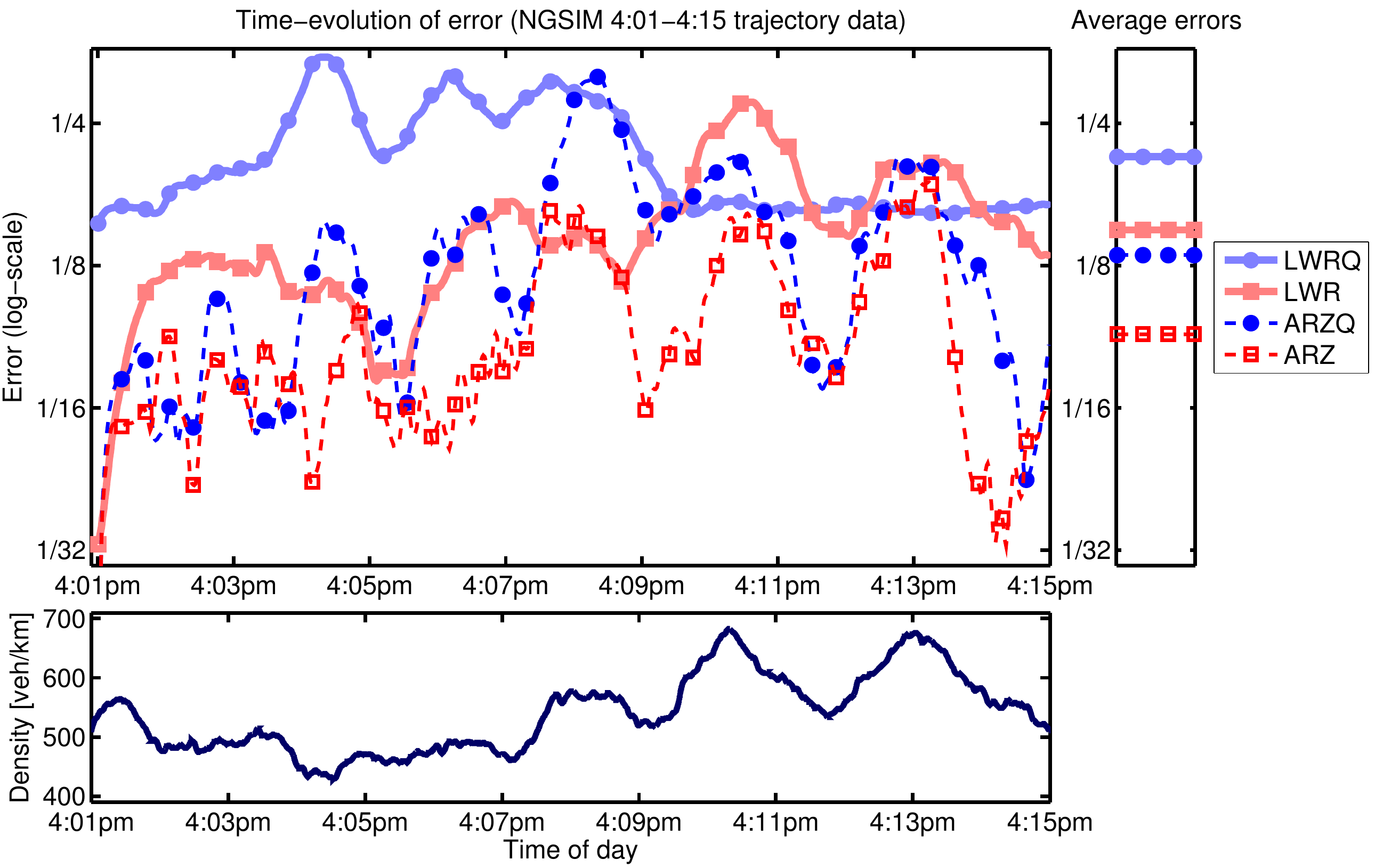}
\vspace{-1.5em}
\caption{Comparison of traffic models for the 4:01--4:15 NGSIM trajectory data set. Shown is the temporal evolution of the errors (left), and the space-time average errors (right), all in a log-scale, as well as the traffic density evolution (bottom).}
\label{fig:results_ngsim}
\end{figure}

The test setup described above is applied independently to each of the three NGSIM data sets. There is one caveat with the vehicle trajectories, namely that during the first minute, not all vehicles on the roadway are recorded. We therefore start each simulation one minute after the data set starts, i.e., the three test cases cover the times 4:01pm--4:15pm, 5:01pm--5:15pm, and 5:16pm--5:30pm. We first take a deeper look at the time segment 4:01pm--4:15pm. In order to obtain an impression of the model error at different times, the temporal evolution of the space-averaged errors \eqref{eq:error_x} for the four models is shown in the left panel of Fig.~\ref{fig:results_ngsim}. Here, the two first order models (LWRQ and LWR) are depicted by thick, solid, light-blue and light-red curves, respectively, and their corresponding second order generalizations (ARZQ and ARZ) are shown by thin, dashed, dark-blue and thin dark-red curves, respectively. The according space-time-averaged errors \eqref{eq:error_xt} are shown in the right panel. Below the error plots, the evolution of the traffic density is shown.

Figure~\ref{fig:results_ngsim} reveals a couple of interesting aspects. Clearly, second order models tend to exhibit more oscillatory errors than first order models. This follows from the theory of hyperbolic conservation laws: in first order models, temporal oscillations in the boundary data are turned into shocks, as they are transported into the computational domain, that then decay towards an uniform profile. In contrast, in the second order ARZQ and ARZ models, oscillations of the data at the left boundary can turn into waves (corresponding to contact discontinuities \cite{AwRascle2000}) that are transported into the domain without decaying towards a uniform profile.

\begin{table}
\begin{tabular}{|ll||lr|lr|lr|lr|}
\hline
Data set\rule{0em}{1.04em}\hspace{-1em} & &
LWRQ\hspace{-1em} & & LWR\hspace{-.5em} & & ARZQ\hspace{-1em} & & ARZ\hspace{-1em} & \\
\hline\hline
NGSIM\rule{0em}{1.04em}
      & 4:01--4:15 & 0.242 &(+230\%) & 0.127 &(+73\%) & 0.126 &(+72\%) & 0.073 & \\
NGSIM & 5:01--5:15 & 0.255 &(+201\%) & 0.115 &(+36\%) & 0.149 &(+76\%) & 0.085 & \\
NGSIM & 5:16--5:30 & 0.209 & (+78\%) & 0.124 & (+6\%) & 0.152 &(+30\%) & 0.117 & \\
\hline
RTMC\rule{0em}{1.04em}
     & congested   & 0.352 &(+110\%) & 0.167 &        & 0.173 & (+4\%) &0.172 &  (+3\%)\\
RTMC & non-cong.   & 0.094 & (+94\%) & 0.069 &(+43\%) & 0.048 &        &0.066 & (+37\%)\\
\hline
\end{tabular}
\vspace{.6em}
\caption{Comparison of traffic models for NGSIM data sets (rows 2--4) and RTMC data, separated into congested and non-congested days (rows 5--6). Given are the space-time averaged errors \eqref{eq:error_xt} for the NGSIM trajectory data, and time-and-day averaged errors \eqref{eq:error_days} for the RTMC sensor data. In parentheses shown are the relative excess errors of the models with respect to the most accurate model (for each particular test).}
\label{tab:results}
\end{table}

\begin{figure}
\includegraphics[width=\textwidth]{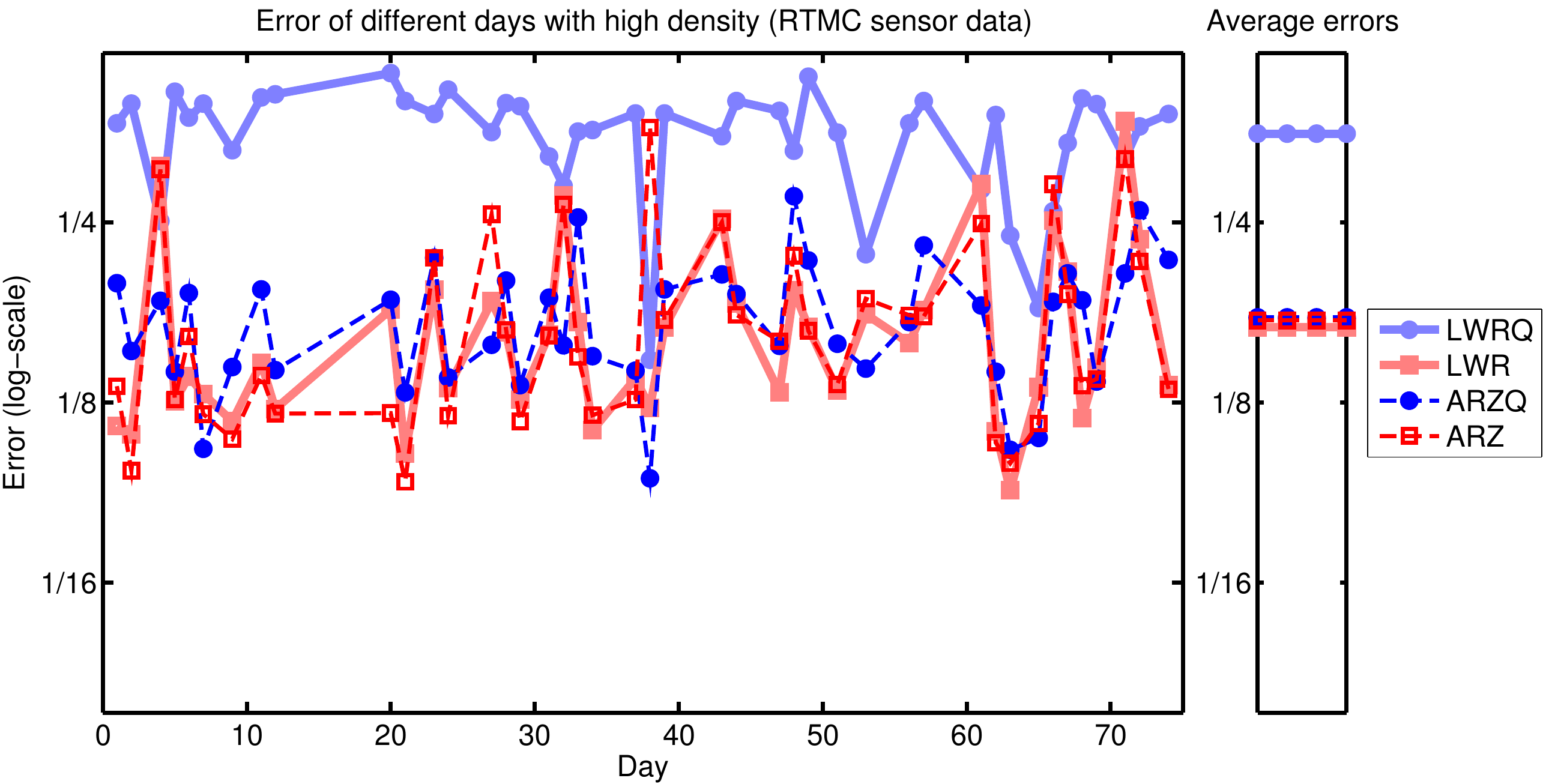} \\[.5em]
\includegraphics[width=\textwidth]{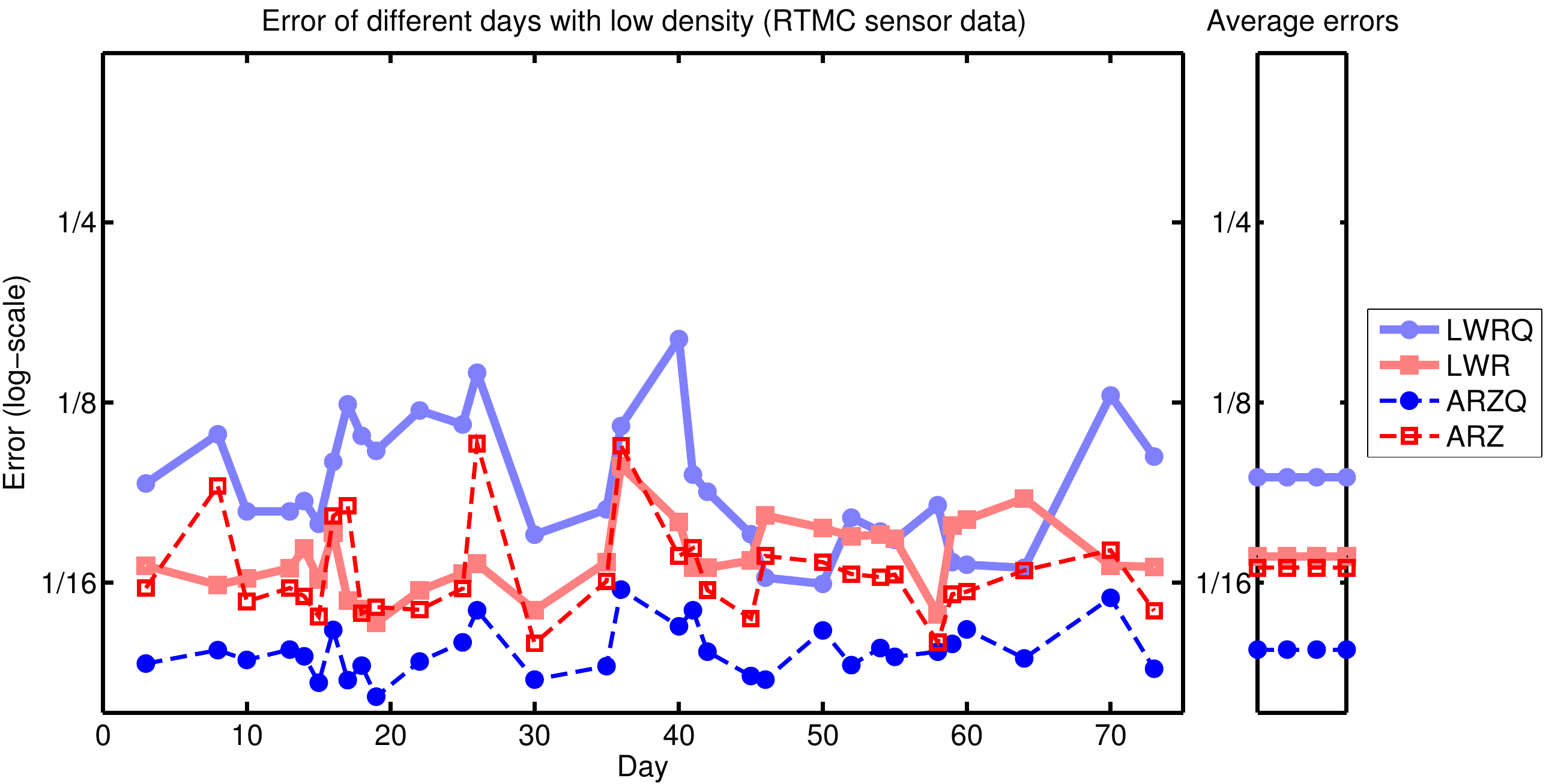}
\vspace{-1.5em}
\caption{Comparison of traffic models for the RTMC sensor data on 74 week days in 2003. The top panels show 43 days with a high congestion level; the bottom panels show the remaining 31 days with low average densities. In each case, the time-averaged errors day-by-day are shown (left), and the effective errors, averaged over the considered days (right). All error plots are in log-scale.}
\label{fig:results_rtmc}
\end{figure}

Another apparent aspect is that even if one model is significantly closer to the data than another model in the space-time averaged sense, this relationship need not hold consistently for all times in the space-averaged errors. For instance, the ARZ model yields a significantly lower total error than the LWR model. However, at a few short instances in time, the LWR solution is closer to the data than the ARZ solution. The same holds true for ARZQ compared to LWRQ. Note that such incidental behavior is not unreasonable in unsteady traffic flow. However, it demonstrates that studies of traffic outside of the free-flow regime require sufficient averaging in order to yield meaningful results.

The main message to be concluded from this test is that the space-time averaged errors of the four models speak relatively clearly in favor of \vspace{-.5em}
\begin{enumerate}[(a)]
\item the three-parameter data-fitted flux function over the Greenshields flux; and
\item second order models over their first order counterparts.
\end{enumerate}
These fully averaged model errors are also given in the first NGSIM row in Table~\ref{tab:results}, together with the relative excess error of the LWRQ, LWR, and ARZQ model to the (smallest) ARZ model error, shown in parentheses. The second and third NGSIM row in Table~\ref{tab:results} then show the full model errors (and the relative excess errors) of the four models for the NGSIM data sets at the other two time segments (5:01pm--5:15pm and 5:16pm--5:30pm). The general trends (a) and (b) described above persist, with one caveat: as the traffic becomes more congested (later time segments), the improvements due to better quality FD curves, and due to using a second order model, are less pronounced than they are in the transitional time segment 4:01pm--4:15pm. The fact that for higher traffic densities, second order models become less advantageous over their first order counterparts, could be related to the unrealistic non-constant stagnation density that the ARZ model possesses (see Fig.~\ref{fig:models}).

\subsection{Model Comparison using the RTMC Sensor Data}
\label{subsec:comparison_rtmc}
The three-detector problem test setup is now applied to the RTMC sensor data, independently for 74 week days, and on each of these days for the hour between 4pm and 5pm. An inspection of the average traffic densities at each day reveals a relatively clear separation between days at which traffic becomes congested, and days at which traffic densities remain moderate. We therefore separate the 74 days into 43 ``congested days'', defined as the time-averaged traffic density (aggregated over the four lanes) exceeding 80 vehicles per km, and ``non-congested days'', which are the remaining 31 days.

Figure~\ref{fig:results_rtmc} shows the time-averaged errors \eqref{eq:error_t} for the 43 congested days (top), and for the 31 non-congested days (bottom). The day-by-day errors are given in the left panels, and the right panels show the errors averaged over the respective list of days. In all panels, the line style conventions are the same as in Fig.~\ref{fig:results_ngsim}. These fully averaged errors are also given in rows 5--6 in Table~\ref{tab:results}, as are the excess errors relative to the best model in parentheses.

The results obtained with the RTMC data are not as conclusive as those obtained with the NGSIM data. Most likely, the significant noise level, measurement errors, and reconstruction errors of the single loop sensors play an important role in this. Still, a variety of conclusions can be drawn from the results. The fact that for the Greenshields flux, the second order model (ARZQ) yields more accurate results than the first order model (LWRQ), persists. In contrast, with the better-fitted three-parameter flux function \eqref{eq:flow_rate_vs_density_function}, the differences between ARZ and LWR are marginal. What can be observed is that there are a few days (e.g., 26 or 38) for which ARZ is much less accurate than LWR, while for the majority of days ARZ is more accurate than LWR (by a smaller margin). An inspection of the solutions on day 38 reveals an explanation for the large error of ARZ: the densities and velocities at the boundaries happen to create wave speeds very close to zero, and as a consequence, boundary data that is fed into the model does not reach sensor 2 within a reasonable time. A surprising result is that for the non-congested days, ARZQ yields smaller errors (factor 1.4) than LWR and ARZ. We have no explanation for this phenomenon.

\vspace{1.5em}
\section{Conclusions and Outlook}
\label{sec:conclusions_and_outlook}
We have established a systematic methodology to compare the predictive accuracy of data-fitted macroscopic traffic models by using time-dependent data of two different types: vehicle trajectories and stationary sensors. The traffic models studied here are first order Lighthill-Whitham-Richards (LWR) models, whose flux functions are obtained via a least squares fit to historic fundamental diagram (FD) data, and their second order Aw-Rascle-Zhang (ARZ) generalizations. The presented approach is a fully macroscopic version of the three-detector problem test \cite{Daganzo1997}, in the sense that the discrete data is transformed into continuously defined field quantities and that discretization effects are negligible. The following conclusions can be drawn from the studies:
\begin{itemize}
\item
The use of better data-fitted flux functions tends to lead to more accurate models. The only exception from this rule is observed for sensor data in a non-congested regime.
\item
Second order models are consistently as good as their first order counterparts, and in many cases they are significantly more accurate. This is remarkable, because the ARZ model does not contain more data-fitted parameters than the LWR model. A key reason for the better quality of second order models is that they allow for more independent information be fed into the solution through the boundaries.
\end{itemize}

This study also gives rise to a variety of further opportunities and questions that should be investigated. Most importantly, the ARZ model's limitation of possessing a non-unique maximum density should be remedied. ``Generic second order'' traffic models \cite{LebacqueMammarHajSalem2007} provide a framework to achieve this goal. We shall consider the predictive accuracy of such types of generalized ARZ models in an upcoming paper \cite{FanHertySeibold2012}.

Another important model extension are inhomogeneous second order models. A simple version is the model \eqref{eq:aw_rascle_zhang_model_inhomogeneous} with $h(\rho) = U(0)-U(\rho)$, whose solutions relax towards LWR solutions. However, as pointed out in \cite{Greenberg2004, FlynnKasimovNaveRosalesSeibold2009}, second order models with relaxation terms can possess much richer dynamics, if one chooses the functions $U(\rho)$ and $h(\rho)$ independently of each other. In this case, the models can develop instabilities (``phantom traffic jams'') and traveling wave solutions (``stop-and-go waves''). Moreover, as shown in \cite{SeiboldFlynnKasimovRosales2012}, traffic waves can explain the observed spread in FD data, and therefore this generalization could give rise to a construction of $U(\rho)$ and $h(\rho)$ in different ways than conducted in this paper.

An alternative class of models, that are specifically designed to match set-valued FDs, are phase-transition models \cite{Colombo2003, BlandinWorkGoatinPiccoliBayen2011}. The incorporation of such models into the comparison framework established here would be an interesting aspect for future research.

A different direction would be the extension of the macroscopic comparison framework to other types of data, such as GPS data, as used in the Mobile Millennium project \cite{WorkTossavainenBlandinBayenIwuchukwuTracton2008}. And finally, it would be of interest to incorporate online calibration techniques, such as Kalman filters, into some of the traffic models studied here, and to investigate to which extent their predictive accuracy can be further improved.

\vspace{1.5em}
\section*{Acknowledgments}
The authors would like to thank Michael Herty, Benedetto Piccoli, and Rodolfo Ruben Rosales for helpful discussions. B. Seibold would like to acknowledge the support by the National Science Foundation though grant DMS--1007967, and partial support through grants DMS--0813648 and DMS--1115269. This research was supported in part by the National Science Foundation through major research instrumentation grant number CNS--09--58854.

\vspace{1.5em}
\bibliographystyle{plain}
\bibliography{references_complete}

\end{document}